\documentclass[conference]{IEEEtran}
\usepackage{url}
\usepackage{subfigure}
\usepackage{graphicx}
\usepackage{lipsum}
\usepackage{adjustbox}
\usepackage{tabularx}
\usepackage{multirow}
\usepackage{enumitem}
\usepackage{amsmath}
\usepackage{framed}
\begin{document}
\title{Characterizing Linguistic Attributes for Automatic Classification of Intent Based Racist/Radicalized Posts on Tumblr Micro-Blogging Website}
\author{\IEEEauthorblockN{Swati Agarwal}
\IEEEauthorblockA{Indraprastha Institute of Information Technology\\
New Delhi, India\\
Email: swatia@iiitd.ac.in}
\and
\IEEEauthorblockN{Ashish Sureka}
\IEEEauthorblockA{ABB Corporate Research\\
Bangalore, India\\
Email: ashish.sureka@in.abb.com}}
\maketitle
\begin{abstract}
Research shows that many like-minded people use popular microblogging websites for posting hateful speech against various religions and race. Automatic identification of racist and hate promoting posts is required for building social media intelligence and security informatics based solutions. However, just keyword spotting based techniques cannot be used to accurately identify the intent of a post. In this paper, we address the challenge of the presence of ambiguity in such posts by identifying the intent of author. We conduct our study on Tumblr microblogging website and develop a cascaded ensemble learning classifier for identifying the posts having racist or radicalized intent. We train our model by identifying various semantic, sentiment and linguistic features from free-form text. Our experimental results shows that the proposed approach is effective and the emotion tone, social tendencies, language cues and personality traits of a narrative are discriminatory features for identifying the racist intent behind a post.
\end{abstract}
\IEEEpeerreviewmaketitle
\section{Introduction}
\begin{figure}[t]
\centering{
\subfigure[\label{fig:example_religion_yes}Topic- Religion, Intent-Yes]{\includegraphics[width=0.48\textwidth]{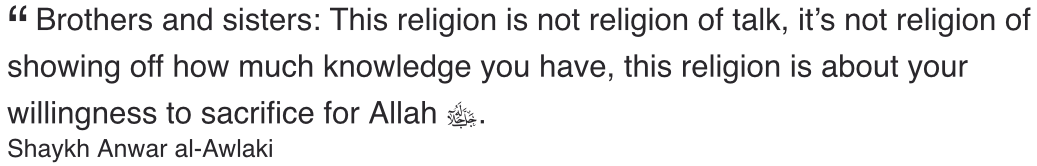}}
\subfigure[\label{fig:example_religion_no}Topic- Religion, Intent- No]{\includegraphics[width=0.48\textwidth]{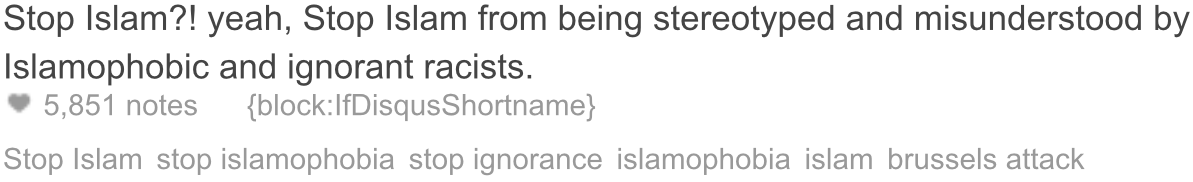}}
\subfigure[\label{fig:example_race_yes}Topic- Race, Intent- Yes]{\includegraphics[width=0.48\textwidth]{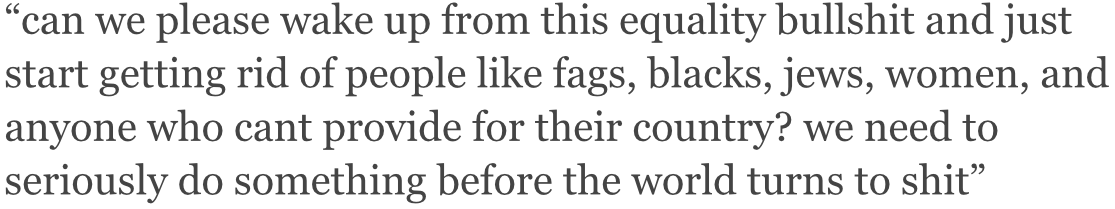}}
\subfigure[\label{fig:example_race_no}Topic- Race, Intent-No]{\includegraphics[width=0.48\textwidth]{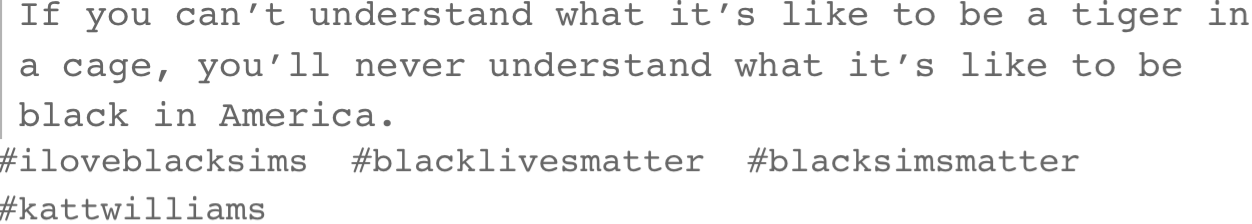}}
}
\caption{Concrete Examples of Tumblr Posts Showing the Different Topics of Racism and Different Intent of Bloggers}\label{fig:examples}

\end{figure}

Freedom of expressions provides leverage to an individual to share their opinions and beliefs about anything. However, many like-minded people misuse freedom of expression to make offensive comments or promote their beliefs that can lead to a negative impact on society \cite{10.2307/2265305}. Research shows that these individuals or groups of people use popular microblogging websites (Twitter and Tumblr) for such activities \cite{smith2008language}\cite{agarwal2015open}. We find that there are users who misuse freedom of speech to post abusive and aggressive comments about a targeted people and other users who promote their beliefs about certain religion or community. Existing literature shows that racism is not specific to only minor communities. There are users who post racist comments targeting existing like-minded groups calling it as reverse racism \cite{norton2011whites}. For example, anti-white bias groups posting comments against white supremacy communities while Islamophobic groups posting hateful speech against Muslim communities. We will be using blogger, author, user and narrative terms interchangeably. Based on our analysis, we broadly define these groups into two categories: Religion and Race. Figure \ref{fig:example_religion_yes} and \ref{fig:example_religion_no} shows examples of two Tumblr posts where bloggers mention about the Islam religion. In Figure \ref{fig:example_religion_yes}, the intention of author is to provoke his Muslim followers for Jihad and develop a willingness to sacrifice themselves for their religion, whereas in Figure \ref{fig:example_religion_no}, the intention of author is to bring awareness that Islamophobic and other hate groups should stop misunderstanding Islam religion. This post was made on March $25$, $2016$ when \#StopIslam hashtag was trending on Twitter. Similarly, Figure \ref{fig:example_race_yes} and \ref{fig:example_race_no} shows examples of two different Tumblr posts where authors talk about black communities. Figure \ref{fig:example_race_yes} depicts that the intent of author is to make a hateful and offensive post targeting various communities. While, in Figure \ref{fig:example_race_no} author's intent is to highlight the challenging life of Black people in America and showing their support for them.\\

In this paper, we conduct our study on Tumblr microblogging website and address the challenge of mining intention of a narrative behind such posts. 
Intent mining from free-form social-media text is technically challenging problem due to the presence of multi-lingual script, incorrect grammar, misspell words, short text, acronyms and abbreviations, sarcasm and opinion based posts. We also find examples of ambiguous content which makes a post difficult to classify even for human annotation. For example, in a Tumblr post "\textit{Yes I enjoyed and actually love CACW but I am so pissed off that the people who suffer / die In the end are a woman (Peggy dies) and two black men (T'challa's father dies and Rhodes is paralyzed) Thank you marvel}", author has posted a movie review with an intention to highlight the racism and target a community of viewers who did not find it racist. However, when we did a manual inspection on the blogger's page; we found that the author is actively involved in bashing and using foul language against certain blogs supporting MCU (Marvel Cinematic Universe) and hence the intent of author is not to support the women or black people but to make negative posts about MCU. Further, it is technically challenging to identify the intent of a post when a naive post has similar terms as a radicalized or racist post. For example, a post P1: \textit{"All types of Jihad is to establish peace for all \& Sharia also promote peace so there is no need to fix anything @simafaysal @profdstone"}- posted by an author with screen name 'Prisoner' and an another post P2: \textit{"This settles it? 'Jihad is to establish peace' 'there is no need to fix anything' spoken like a true 'prisoner'"} have similar content. Here, the intention of P1 is to show his support for Jihad and terrorism while intention of P2 is to make a sarcastic comment on P1 and author's belief. Further, despite having hateful comments in a post, the intention of author can still be naive. For example, in January $2016$, Saudi Arabia released an official video on 'how to properly beat Muslim women' with an intention of targeting women communities. Recently, as the video got published worldwide, users at microblogging websites shared that video and posted hateful comments in order to oppose the video with no racist intent. Whereas, some users posted comments opposing the video and targeting whole Muslim community with racist intentions bringing ambiguity in their posts. Tumblr website is popularly known for the use of gif images where users share their opinions by embedding reaction gifs in their posts. It also allows users to share content from external sources such as news websites or blogs (wordpress). Users can disguise themselves by sharing only articles or external URLs in their posts. Therefore, automatic identification of narrative's intentions in such posts is a significantly technically challenging problem. \\

The work presented in this paper is motivated by the need to develop a system for automatically identifying the intent of a racist and radicalized post. The specific research aim of the study presented in this paper is the following:
\begin{enumerate}[topsep=0pt]
 \item To investigate the efficacy of natural language processing techniques on microblogging dataset for topic and intent classification.
 \item To investigate the application of linguistic features such as taxonomy, emotions, language cues, personality traits and text semantics for classifying the intent of Tumblr posts.
 \item To conduct empirical analysis on real word dataset and examine the effectiveness of proposed one-class text classification approach. To compute the relative influence of each linguistic feature for identifying the posts having racist intent.
\end{enumerate}
\section{Related Work}
In this Section, we discuss closely related work to the study presented in this paper. We conduct a literature survey in the area of intent mining on social media platforms and divide our related work into following two categories:\\
\indent \textbf{Commercial Intent Classification: }Wang et al. \cite{wang2015mining} present a graph based semi-supervised learning technique to classify intent tweets. They combine keyword based flagging (referred as intent keyword) and graph regularization method for classifying tweets into six categories. Purohit et al. \cite{purohit2015} present an hybrid approach of combining knowledge-guided patterns and bag-of-tokens model for intent classification of short text. They conduct a study on Twitter for crisis events dataset and address the problem of ambiguity and sparsity in order to classify the intent of narrative. Ding et al. \cite{ding2015mining} present a transfer learning based convolutional neural network model for identifying users' buying or consumption intentions from Sina Weibo- a Chinese microblogging service\footnote{\url{http://weibo.com/}}. Geetha et al. \cite{geetha2014feature} present a lexicon (sentiment Wordnet dictionary) based bootstrapping method to measure the polarity of opinion in short text data. They conduct a study on Twitter data and compare their results for movie reviews, election results and product reviews. Wang et al. \cite{wang2013mining} present a graph based ranking model to identify the commercial intent from trending topics on microblogging platforms.\\
\indent \textbf{Racism/Radicalization Intent Classification: }Smith et al. \cite{smith2008language} conduct a quantitative content analysis on public documents to distinguish radical groups from non-radical groups. Prentice et al. \cite{prentice2011analyzing} conduct a quantitative text analysis on $50$ documents originated from extremist websites. They present a 'Conduct and Composition Analysis' technique to classify the persuasion behavior of online extremist media varying for the documents posted before and after the Israeli activities in Gaza. Our literature survey reveals that there has been a lot of work in the area of commercial intention identification from free-form text whereas automatic detection of racist posts on social media platforms such as Tumblr is a relatively unexplored area. 
\section{Research Contributions}
In contrast to the existing work, our paper makes the following novel contributions:
\begin{enumerate}[topsep=0pt]
 \item To the best of our knowledge, the study presented in this paper is the first work on racist and radicalization detection based on the intent of narrative unlike previous keyword spotting methods.
 \item We apply natural language processing techniques on Tumblr posts for identifying discriminatory features for intent classification.
 \item We publish the first ever semantically and sentimental enriched data of Tumblr posts and make our data publicly available for benchmarking and extension\footnote{\url{http://dx.doi.org/10.17632/hd3b6v659v.2}} \cite{agarwaltumblr}.
 \item The study presented in this paper is an extended version of our work Agarwal et al. accepted as a short paper in European Intelligence and Security Informatics Conference (EISIC $2016$) \cite{agarwal2016eisic}. Due to the small page limit for short papers (at most four pages) in EISIC $2016$\footnote{\url{http://eisic.org/eisic2016/}}, several aspects including results and details of proposed approach are not covered. This paper presents the complete and detailed description of our work on intent based classification of racist and radicalized posts made on Tumblr micro-blogging website.

\end{enumerate}
\section{Problem Statement}
Given a dataset $D$ of Tumblr Posts $P_{i}$, $D=\{P_{i} \mid 1 \le i \le n \}$, a set of topics $N=\{N_{j} \mid 1 \le j \le m \}$ and a target class $C$; identify the intent of $P_{i} \in D$ when $P_{i} \in N$.

Based on the definition of freedom of expression by Joshua Cohen \cite{10.2307/2265305}, we define a Tumblr post $P_{i}$ as a racist intent post if 1) the topic of the content belongs to a race or a religion and 2) the post targets a community in an offensive or persuasive manner (in a recognizable way). In order to identify a racist or radicalized intent post, we propose following two hypotheses:
\begin{enumerate}
 \item In the absence of topic related key-terms, natural language processing can be an efficient approach to identify hidden taxonomy of a Tumblr post.
 \item Sentiment and semantic enrichment of text can be two discriminatory features for identifying the language of narrative and classifying the intent posts.
\end{enumerate}
\section{Experimental Setup}\label{sec:experimental_setup}
\textbf{Data Collection: }We conduct our experiments on an open source and real time dataset extracted from Tumblr microblogging website. We perform a manual inspection and find most popular Tumblr posts having racist and radicalized intent. We extract the list of unique tags associated with these posts and create a lexicon of top \textit{K} tags that are the most commonly used by racist or radicalized groups. For example, \#islamophobia, \#islam is evil, \#supremacy, \#blacklivesmatter, \#white racism, \#jihad, \#isis and \#white genocide. We implement a bootstrapping method to create our dataset and use this lexicon as seed tags for the Tumblr Search API\footnote{\url{https://www.tumblr.com/docs/en/api/v2}}. For each tag, we extract only textual posts (text and quote) and extend our lexicon by acquiring other (unique) related tags associated with these posts. We execute our model until we get a desired number of posts or the model converges (it starts extracting duplicate posts). Using Tumblr Search API, we were able to extract a total of $3,228$ text posts made by $2,224$ unique bloggers consisting of $10,217$ unique tags. Table \ref{tab:schema} shows a complete schema of additional metadata extracted for each post and unique blogger. The aim of the study presented in this paper is to build a one-class text classifier for identifying racist and hate promoting intent posts. Therefore, we conduct our experiments on post content (referred as description in Tumblr). Since, Tumblr generates a new identification number for each post (re-blogged or posted), despite having the unique Post IDs, we discard $9$\% ($273$) of the posts having similar or duplicate content and remove the bias from our data.\\
\begin{table}[t]
\normalsize
\sffamily
 \centering
 \caption{\textbf{Detailed Schema of Tumblr Database Consisting of Posts and Bloggers' Metadata}}
 \begin{tabular}{|p{0.46\textwidth}|}
 \hline
 \multicolumn{1}{|c|}{\textbf{Posts}}\\
 Post\_ID $\mid$ Timestamp $\mid$ GMT $\mid$ Blogger $\mid$ URL $\mid$ Type $\mid$ Tags $\mid$ Num\_Tags $\mid$ Notes $\mid$ Re-Blogged\_From $\mid$ Title $\mid$ Description\\
 \multicolumn{1}{|c|}{\textbf{Blogger}}\\
 Blogger\_ID $\mid$ Ask $\mid$ Ask\_Anon $\mid$ \#Likes $\mid$ \#Posts $\mid$ Title $\mid$ Description\\
 \hline
 \end{tabular}
 \label{tab:schema}
\end{table}
\begin{figure}
 \centering
 \includegraphics[width=0.48\textwidth]{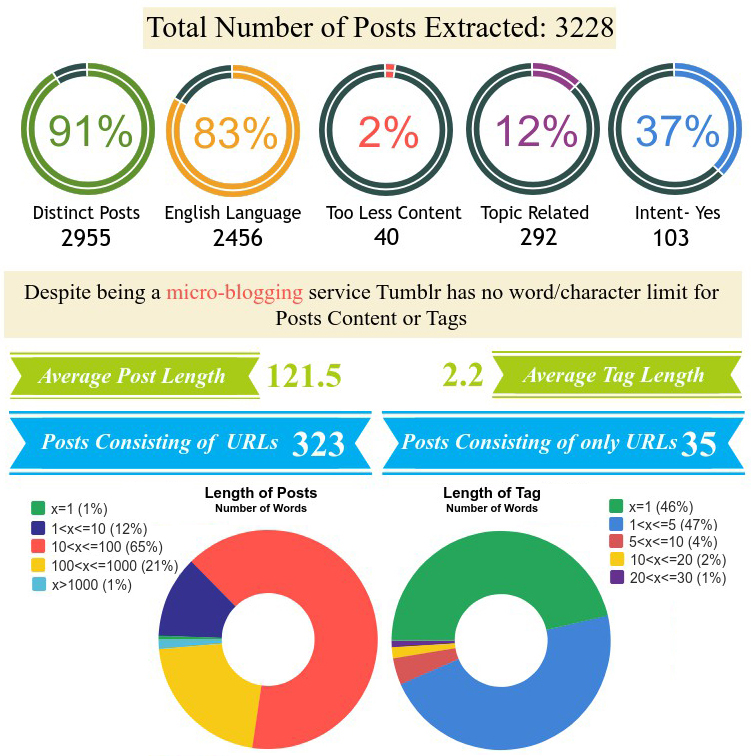}
 \caption{\textbf{Basic Statistics of English Language Posts from Experimental Dataset}}
 \label{fig:stats_infographic}
\end{figure}
\begin{table*}[t]
\normalsize
\sffamily
\caption{\textbf{Inter-Annotator Agreement Results for Topic and Intent Labelling of Experimental Dataset}. Source: Agarwal et al. \cite{agarwal2016eisic}}\label{tab:inter_annotator_agreement}
\centering
\subtable[\label{topic_annotation}Topic Annotation]{%
\begin{tabular}{ll|ll}
\multicolumn{2}{c}{} &\multicolumn{2}{c}{A2}\\
\multicolumn{2}{c|}{} &Topic& NA\\
\cline{2-4}
\multirow{2}{*}{A1}& \multirow{1}{*}{Topic} & 292 & 24\\
& \multirow{1}{*}{NA} & 13 & 2127\\
\end{tabular}}%
\hfill
\subtable[\label{intent_annotation}Intent Annotation]{%
\begin{tabular}{ll|ll}
\multicolumn{2}{c}{} &\multicolumn{2}{c}{A2}\\
\multicolumn{2}{c|}{} &Intent& NA\\
\cline{2-4}
\multirow{2}{*}{A1}& \multirow{1}{*}{Intent} & 103 & 2\\
& \multirow{1}{*}{NA} & 12 & 175\\
\end{tabular}}%
\hfill
\subtable[\label{cohen_kappa}Cohen's Kappa Coefficient]{%
\begin{tabular}{l|ll}
\multicolumn{1}{c|}{} & Topic & Intent\\
\hline
Observed Agreement Po & 0.98 & 0.95\\
Random Agreement Pr(e) & 0.77 & 0.51\\
Kappa Coefficient $\kappa$ & 0.91 & 0.95\\
\end{tabular}}%
\end{table*}
\begin{figure*}[t]
 \centering
 \includegraphics[width=0.75\textwidth]{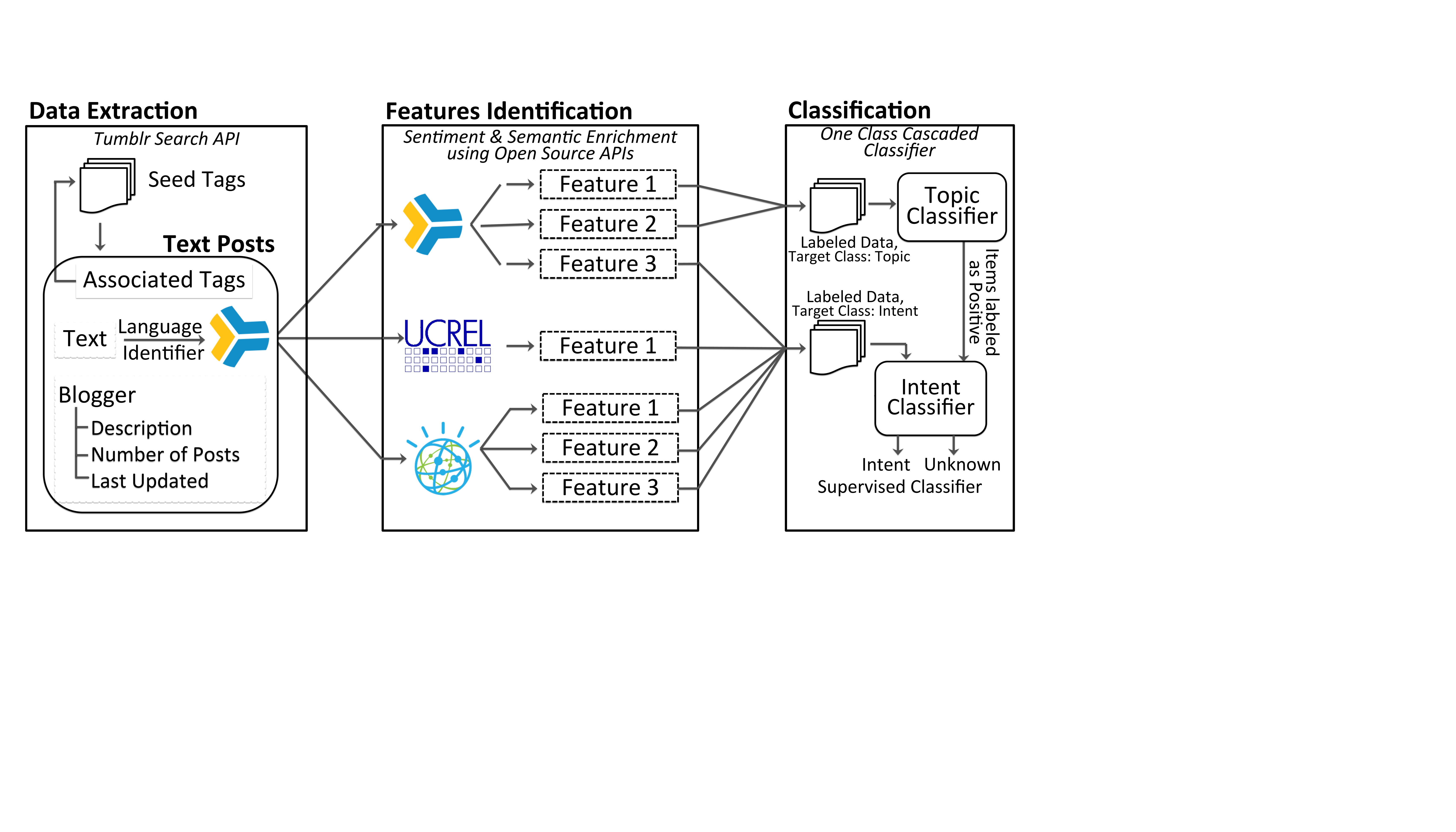}
 \caption{\textbf{A General Research Framework for the Experimental Setup and Proposed Methodology}}
 \label{fig:framework}
\end{figure*}

The study presented in this paper focuses on intent mining on English language posts. We identify the language of each record by applying Alchemy language detection API\footnote{\url{http://www.alchemyapi.com/api/language-detection}} on post description. Figure \ref{fig:stats_infographic} reveals that only $83$\% ($2,456$ out of $2,955$) of the posts have English language content and $459$ posts were identified as non-English. The language of remaining $40$ posts ($2$\% of the data) was identified as 'unknown' due to the insufficient content in post description, for example, the posts containing only URLs. Figure \ref{fig:stats_infographic} reveals that $35$ out of $2,955$ posts contain only URLs. We conduct our experiments on $2,456$ English language posts and discard the other non-English or unknown language records. We apply various natural language processing techniques for semantic and sentiment enrichment of our data (discussed in Section \ref{subsec:features}). We enhance our data and make it publicly available so that our experiments can be used for benchmarking and comparison \cite{agarwaltumblr}. Our dataset is the first ever published data of Tumblr posts and bloggers labeled with various sentiment and semantic features and can be downloaded from Mendeley Data\footnote{\url{https://data.mendeley.com/datasets/hd3b6v659v/2}}. Figure \ref{fig:stats_infographic} summarizes the statistics of our experimental dataset. Despite being a microblogging website, Tumblr has no word or character limit and allows users to make long posts and tag with any number or length of keywords. We remove all noisy text from the post descriptions and tags including special characters, emoticons, extra white spaces and compute their length. Data statistics reveals that $21$\% of the posts have word length between $100$ and $1,000$ while $25$ posts have length greater than $1,000$ words. Similarly, $4$\% ($408$ out of $10,217$) of unique tags have a word length between $5$ to $10$ while $10$ unique tags have a length between $20$ to $30$ words.\\

\textbf{Data Annotation: }
We use $2456$ English language posts for annotation which spans only $83$\% of the extracted data. Since, we are using bootstrapping method to collect our data, it extracts a large number of noisy posts that do not belong to the defined topic (race and religion). Therefore, we first identify the topic related posts and later label them as intent (racist/radicalized) or unknown (we don't know the intent of the author). To annotate these posts, we employ two annotators with $2$ to $3$ years of experience of using Tumblr website. Each annotator first labels a post as topic or unknown (NA) based on the content description and the tags associated with the post. If a post is annotated as topic then these annotators further label it as intent or unknown (NA). To create ground truth for our data, we measure the inter-annotator agreement and compute Cohen's Kappa coefficient between both annotations.

Table \ref{tab:inter_annotator_agreement} shows the results of topic and intent annotation performed on $2,456$ posts. Table \ref{topic_annotation} reveals that we get $2,419$ ($292$ topic and $2,127$ unknown) posts as same label from both the annotators. We discard the remaining $37$ posts with inconsistent annotation. Both the annotators further label these $292$ topic posts as intent or unknown. Table \ref{intent_annotation} reveals that the annotators agree on $278$ posts ($103$ intent, $175$ unknown) while there is an inconsistency in remaining $14$ posts. Table \ref{cohen_kappa} shows the value of Cohen's Kappa coefficient between annotators for both topic and intent annotation. Results reveal that the annotators agree more than $90$\% of the time. Figure \ref{fig:stats_infographic} shows that the intent posts are only $37$\% of topic posts and only $4$\% of the complete experimental dataset, revealing that the labeled data is highly imbalanced. Since, we use a tag search based bootstrapping method, we analyze all the tags extracted during the process and find that it happens due to the various limitations of user generated tags. For example, presence of noisy content (spell errors), long text, multi-lingual tags, use of featured tags and tags that redirects to a non-topic based post such as 'vote', 'lol', 'media', 'news', 'life', 'travel'.
\section{Proposed Approach}
Figure \ref{fig:framework} shows the high-level architecture of proposed approach primarily consisting of three phases: Data Extraction, Feature Identification and Classification. Section \ref{sec:experimental_setup} describes the bootstrapping method used for data collection and inter-annotator agreement used for creating ground truth. We describe the remaining two phases in the following sections:
\subsection{Features Identification}\label{subsec:features}
Based on the prior literature and our hypothesis design, we create our feature space by analyzing the linguistic features (semantic and sentiment tone) of Tumblr posts. We divide our features set into three categories: Topic Modeling, Tone Analysis and Semantic Tagging. We also discuss other contextual metadata features that can be extracted from Tumblr posts but are not applicable in intent classification.\\

\textbf{Topic Modeling: }The existing literature shows that there has been a lot of work in the area of mining user generated content on social media related to offensive speech \cite{chen2012detecting}, racism and radicalization \cite{agarwal2016spider}\cite{burnap2016us}. However, our analysis and annotation reveals that despite not having certain topic specific key-terms, a post can be an intent post for which keyword based classification method do not work accurately and generates a large number of false alarms \cite{agarwal2015open}. Therefore, we use statistical and natural language processing techniques to perform topic modeling on Tumblr posts. We use Alchemy Taxonomy API\footnote{\url{http://www.alchemyapi.com/products/alchemylanguage/taxonomy}} to classify the post into the most likely topic and sub-topic categories. Alchemy API supports over $1000$ categories broadly divided into $23$ topics. Sub-topic categories allows us to identify the more focused and targeted topic of post (upto $5$ levels of hierarchy). For example, society/crime/personal offense/hate crime. We also use Alchemy Concept Tagging API\footnote{\url{http://www.alchemyapi.com/api/concept-tagging}} to identify the hidden concepts in the text that are similar to human annotation. Alchemy API learns about a post from $9$ linked data resources\footnote{\url{http://www.alchemyapi.com/api/concept/ldata.html}} such as freebase, dbpedia, yago and tags the concepts that are high likely to be related to the given text. For example, for a Tumblr post "\textit{If the Arabs put down their weapons today, there would be no more violence. If the Jews put down their weapons today, there would be no more Israel.}", Alchemy tags "Ashkenazi Jews", "Palestinian people" and "Jewish ethnic divisions" with a confidence score of $0.74$, $0.78$ and $0.70$ respectively. We use these concepts to perform the topic modeling of a text along with the taxonomy. Statistically, the API returns a confidence score of each taxonomy conveying how likely the post belongs to derived category. We discard a category from taxonomy and concept lists if the confidence score is below $40$\%.\\

\textbf{Sentiment and Tone Analysis: } Inspired by the prior literature \cite{prentice2011analyzing}, we investigate language of narrative by analyzing various types of sentiments and personality traits in a post such as document sentiment, social tone, writing tone and emotions. We use Alchemy Document Sentiment API\footnote{\url{http://www.alchemyapi.com/api/sentiment-analysis}} to identify the document-level polarity of overall sentiment of a post. We define five categories of sentiment polarity: strongly negative, negative, neutral, positive and strongly positive 
and categorize each post based on it's sentiment score. The sentiment of a document or post differs from the tone analysis of the content. Sentiment analysis can only identify the positive and negative polarity of a post while tone analysis measures the level of three categories including emotion, social and writing tones. We conduct a linguistic analysis on Tumblr posts using IBM Watson Tone Analyzer API\footnote{\url{https://tone-analyzer-demo.mybluemix.net}}. 
\begin{figure}[t]
 \centering
 \includegraphics[width=0.48\textwidth]{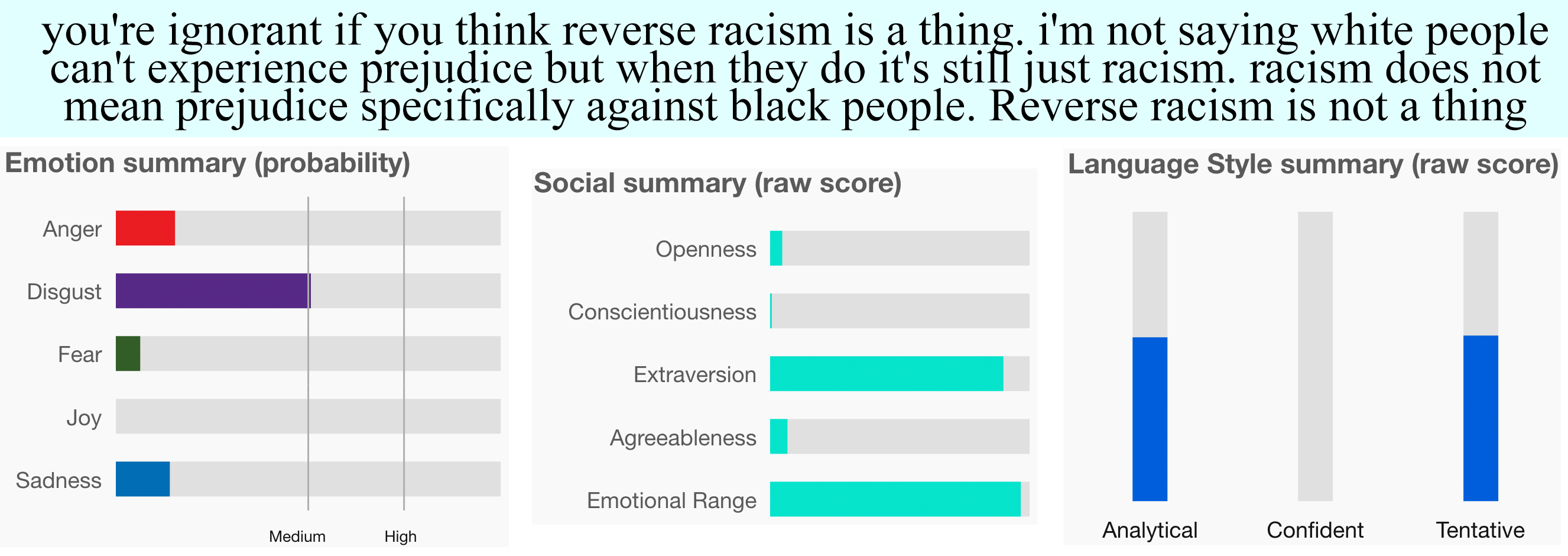}
 \caption{\textbf{Example of Emotion, Social and Writing Tone Features Computed for a Tumblr Post, Topic: Race, Intent: No}}
 \label{fig:tone_example}
\end{figure}
Emotions tone analyzes the text of a post and gives a distribution of $5$ emotions namely joy, fear, sadness, anger and disgust. Social tendencies analyze the personality traits from the text that includes openness, conscientiousness, extraversion, agreeableness and emotional range of a narrative. Writing tone identifies the language cues of the author in context to the content written in a Tumblr post. It includes analytical, confident and tentative style of writing. The Tone Analyzer API analyzes the content of a post and computes two scores (document level and sentence level) for all three categories of tones. Since, the text length of posts in our experimental dataset varies from $1$ to ~$1200$ words, we select only document level measures of these tones. Similar to sentiment score, we create a feature vector of each tone and categorize each post based on the confidence score: very low, low, medium, high, and very high. 
Figure \ref{fig:tone_example} shows a concrete example of Tumblr post related to Race topic and shows the level of emotion tone, language and personality traits of author.\\

\textbf{Semantic Tagging: }Semantic tagging of a post identifies the semantic role of each term present in the content. It also identifies the hidden phrases playing major role in the post. We use UCREL Semantic Analysis System (USAS)\footnote{\url{http://ucrel.lancs.ac.uk/usas/}} to semantically tag each post in our dataset. USAS contains a hierarchy based lexicon of $232$ categories with $21$ major labels at top of the hierarchy. All the semantic tags in a post are composed of a general or high level label and a numeric value showing the division of each label in lexicon. A numeric value after the decimal shows a further sub-division of categories in the hierarchy. For example, term "\textit{refugee}" is tagged as "\textit{M1/S2mf}" where \textit{M1} denotes the tag 'moving from one location to another', \textit{S2} denotes 'people' and \textit{mf} denotes the 'gender'. We use USAS for semantic tagging because it not only tags each word of the document but also tags multi-words unit in the post, if any. For example, term "New York Times" is tagged as "New\_Z3c[i4.3.1 York\_Z3c[i4.3.2 Times\_Z3c[i4.3.3" where \textit{Z3} denotes the name of a company, \textit{c} denotes an anaphora, \textit{i} denotes a multi-words unit and following numeric terms present the number of words present in a unit ($3$). We remove all punctuations and special characters (tagged as PUNC) from semantically tagged content and decode all remaining terms with their respective labels in tags' hierarchy\footnote{USAS published list of all semantic tags is available at \url{http://ucrel.lancs.ac.uk/usas/semtags.txt}}. USAS tags a term as \textit{Z99}, if the term is not identified and not present in USAS database. We however do not remove them from the tagged c	ontent. Because USAS labels various topic specific terms as \textit{Z99} that are important for the intent identification. For example, 'Jihadist', 'racial', 'anti-white', 'pro-black', join words such as 'BlackLivesStillMatter'. It also includes the terms with hashtags, URLs, misspell words, acronyms and abbreviations.\\

\textbf{Contextual Metadata: }Tumblr API allows us to extract the following contextual metadata associated with each Tumblr post: number of tags, terms used in the tags, number of notes (reblog + like count) and link to multimedia content such as image, video or audio attached with the post. By further mining the content of a post, we can extract the following contextual information: hashtags, URLs, emoticons and Internet slang. However, due to various limitations, we exclude these contextual metadata from our feature space. 1) As discussed in Section \ref{sec:experimental_setup}, the length of unique tags present in our dataset varies from $1$ to $35$ and contains a large amount of noisy text (multi-lingual terms, misspell words). Tags are user generated content and a Tumblr post can have any number of tags (upto $30$ in our dataset) or no tags at all. Further, the presence of a comma in a long sentence splits a tag into two separate terms. Given the length of tags in our dataset, number of tags cannot be a discriminatory feature. 2) For a given tag, Tumblr API allows us to extract only most recently published posts. These posts automatically has relatively less number of reblog or like count (referred as notes) in comparison to the posts containing featured or popular tags or uploaded before the current timestamp. Hence, the number of notes is not a valid feature for our experimental data. 3) Since, we extract only textual posts for our analysis, our dataset does not contain any multimedia content such as image, video or audio attached in the post description. 4) We conduct an exploratory data analysis on all topic related posts and our data reveals that for both intent and unknown posts, there are very few (upto maximum $10$) posts that contain either of hashtags (hashtags in Tumblr posts are not clickable and searchable), emoticons, Internet slangs (usually present in tags than the post content), @user mention or external URLs. We exclude contextual metadata from our feature space as those are not discriminatory for intent or topic classification.
\subsection{Classification}\label{subsec:classification}
The third phase of our proposed framework is a cascaded ensemble learning based classifier primarily consisting of two stages: topic classification and intent classification. We train our model from feature vectors created in Phase 2 and perform one-class classification on Tumblr posts.\\

\textbf{Topic Classification: }To identify the posts that belongs to a defined topic (Race or Religion), we use topic modeling linguistic features extracted using natural language processing. We take a random sample of $50$ posts out of $292$, annotated as topic posts and extract their taxonomy and concepts from the feature space. We create two independent lexicons of these concepts and labeled topics that has a confidence score above $0.40$. We manually filter the list of taxonomy and finalize the following $6$ labels that strictly belong to the topic of this study: religion and spirituality, society/unrest and war, society/racism, society/personal offense/hate crime, law, govt \& politics/espionage and intelligence/terrorism and law, govt \& politics/legal issues/human rights.\\ 
\indent We use a look-up based method and check if the post belongs to any of these taxonomies and has a confidence score above $0.40$. If yes, then we classify it as a topic post. However, if a post contains a wide range of taxonomies (\textgreater $5$) then we identify the top \textit{K} concepts in the text and check if they exist in the concept lexicon of labeled topic posts. This stage of cascaded classifier is a one-class classifier that takes complete experimental dataset as an input and classifies topic related posts from unknown posts.\\
\begin{table}[t]
\sffamily
\normalsize
\centering
\caption{\textbf{Feature Codes and Grouping of Similar Feature Vectors}. Source: Agarwal et al. \cite{agarwal2016eisic}}\label{tab:grouped_features}
\begin{tabular}{lp{0.37\textwidth}}
Code & Grouped Features\\
\hline
F1 & Document Sentiment\\
F2 & Semantic Tagging\\
F3 & Emotion \{Anger, Fear, Joy, Disgust, Sadness\}\\
F4 & Writing \{Analytical, Confident, Tentative\}\\
F5 & Social \{Openness, Conscientiousness, Extraversion, Agreeableness and Emotional Range\}\\
\end{tabular}
\end{table}
\begin{figure*}[t]
\centering{
\subfigure[\label{fig:leave1out_dt}Decision Tree]{\includegraphics[width=0.32\textwidth, height=3.05cm]{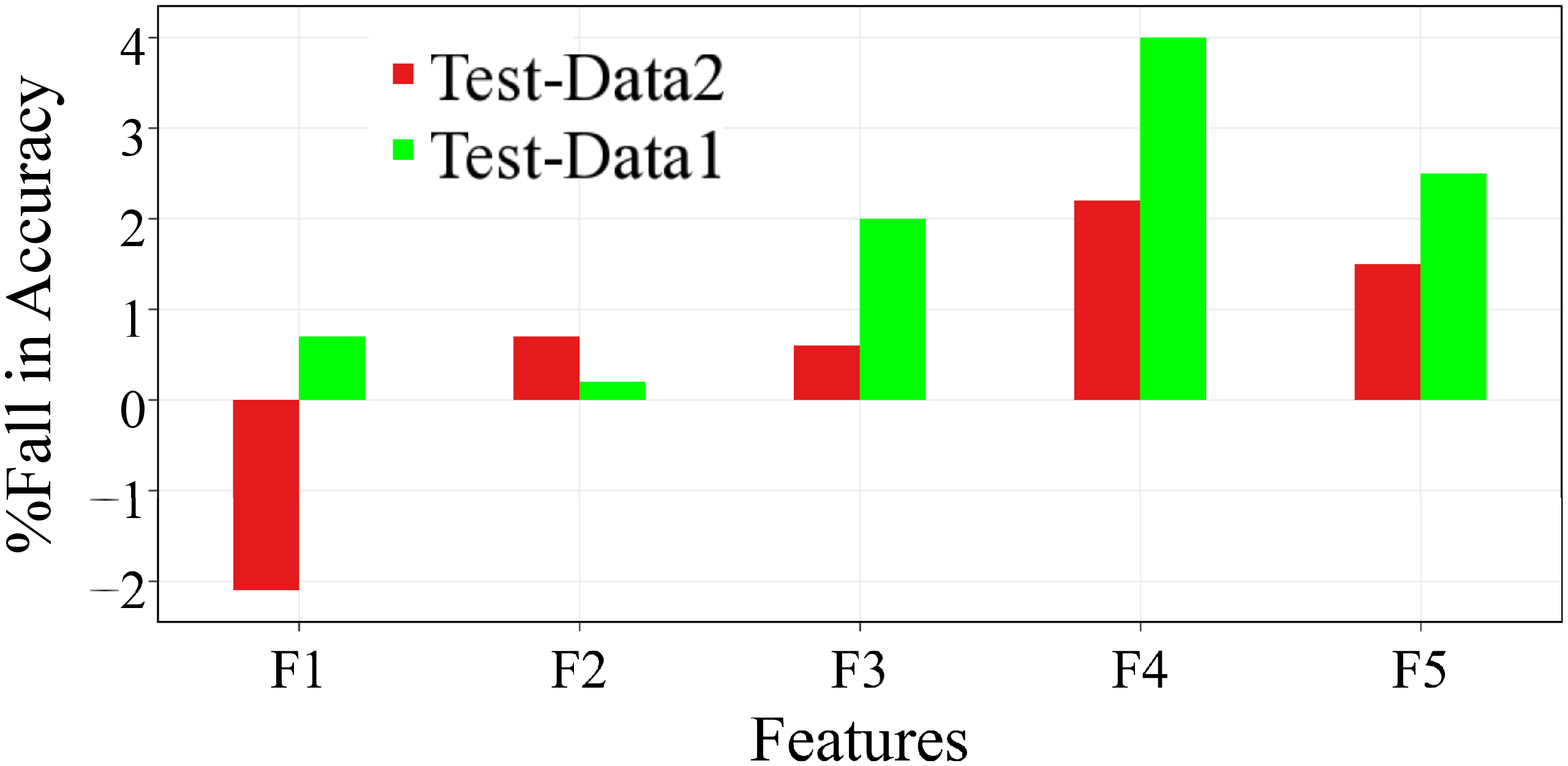}}
\subfigure[\label{fig:leave1out_nb}Naive Bayes]{\includegraphics[width=0.32\textwidth, height=3.05cm]{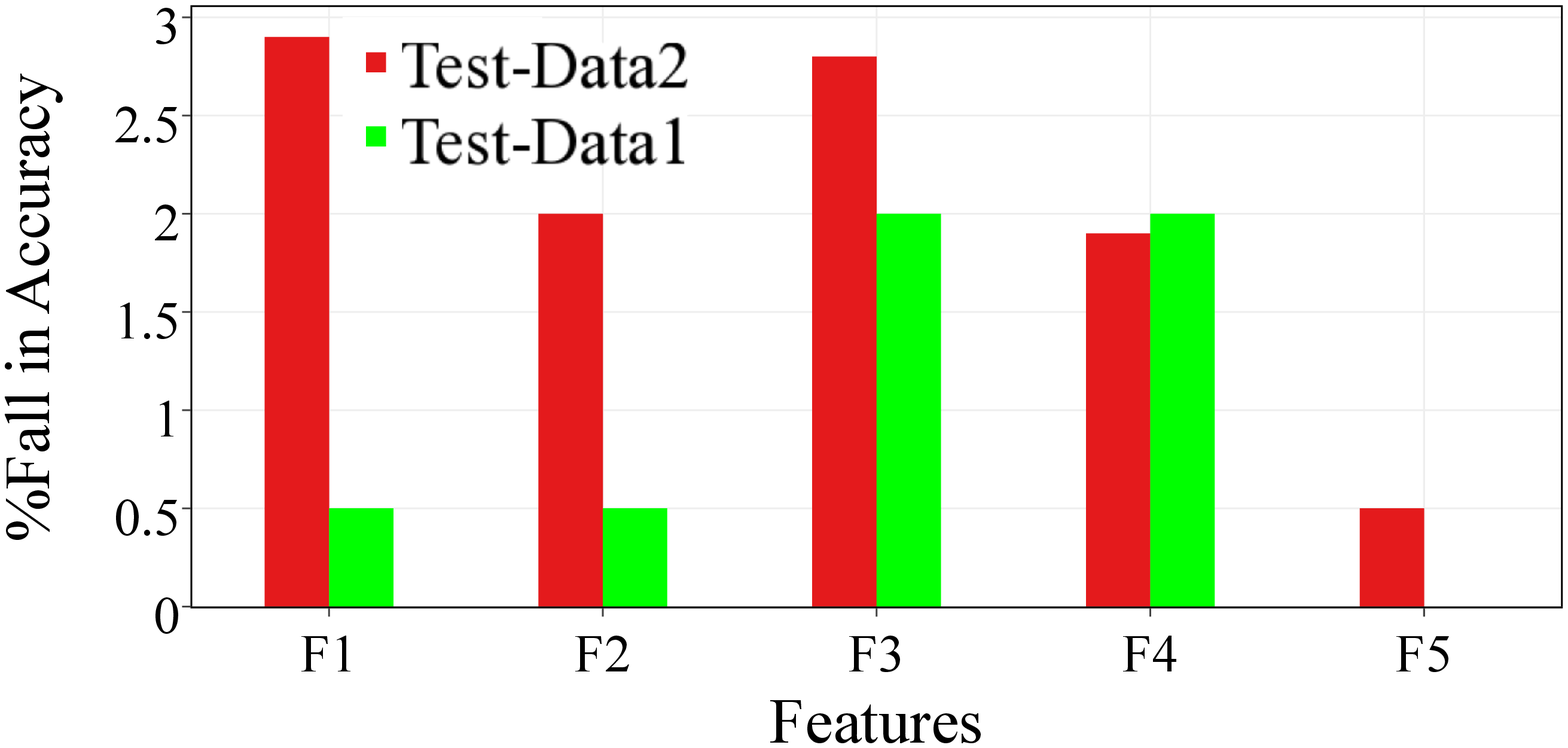}}
\subfigure[\label{fig:leave1out_rf}Random Forest]{\includegraphics[width=0.32\textwidth]{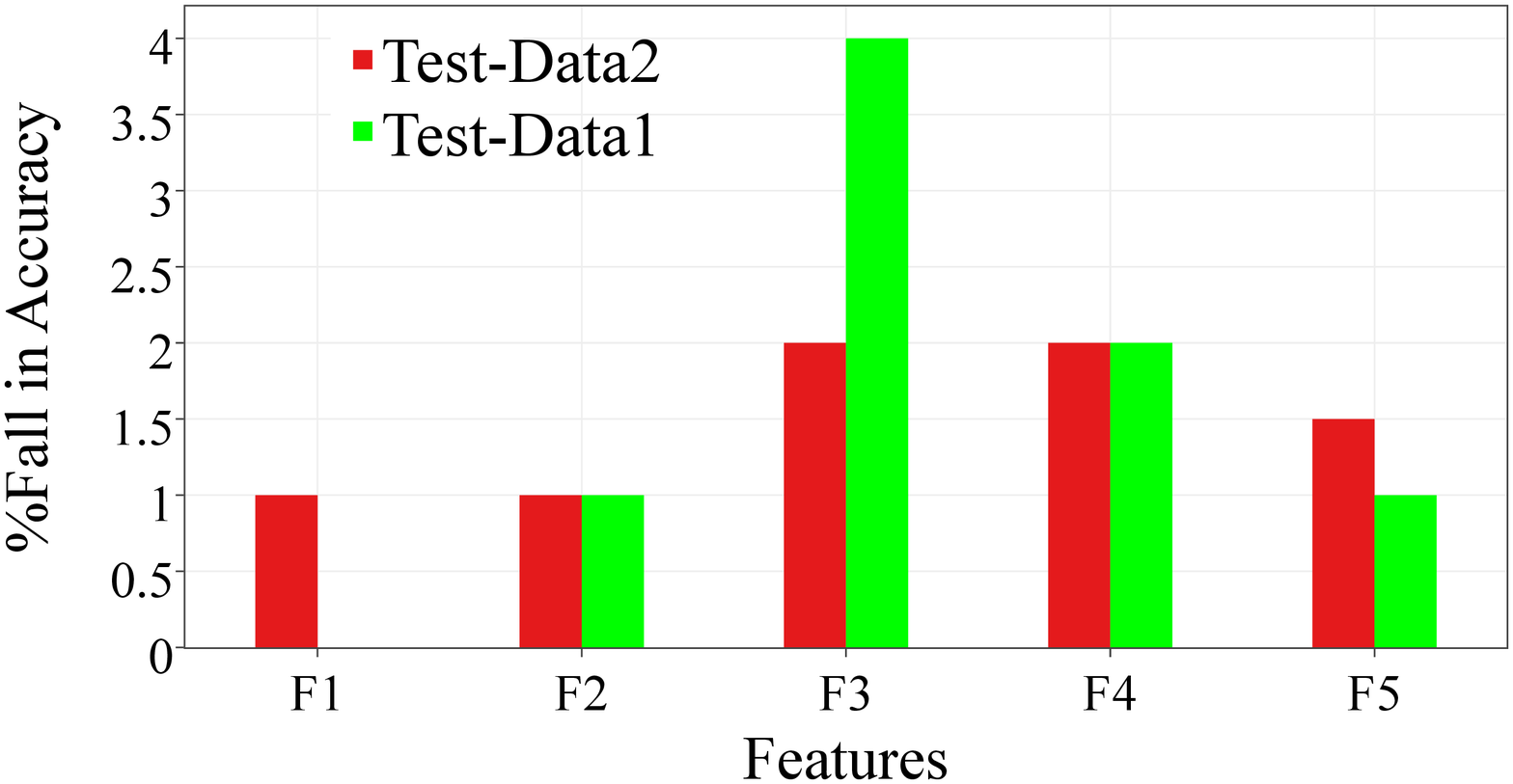}}
\subfigure[\label{fig:leave2out_dt}Decision Tree]{\includegraphics[width=0.32\textwidth, height=3.4cm]{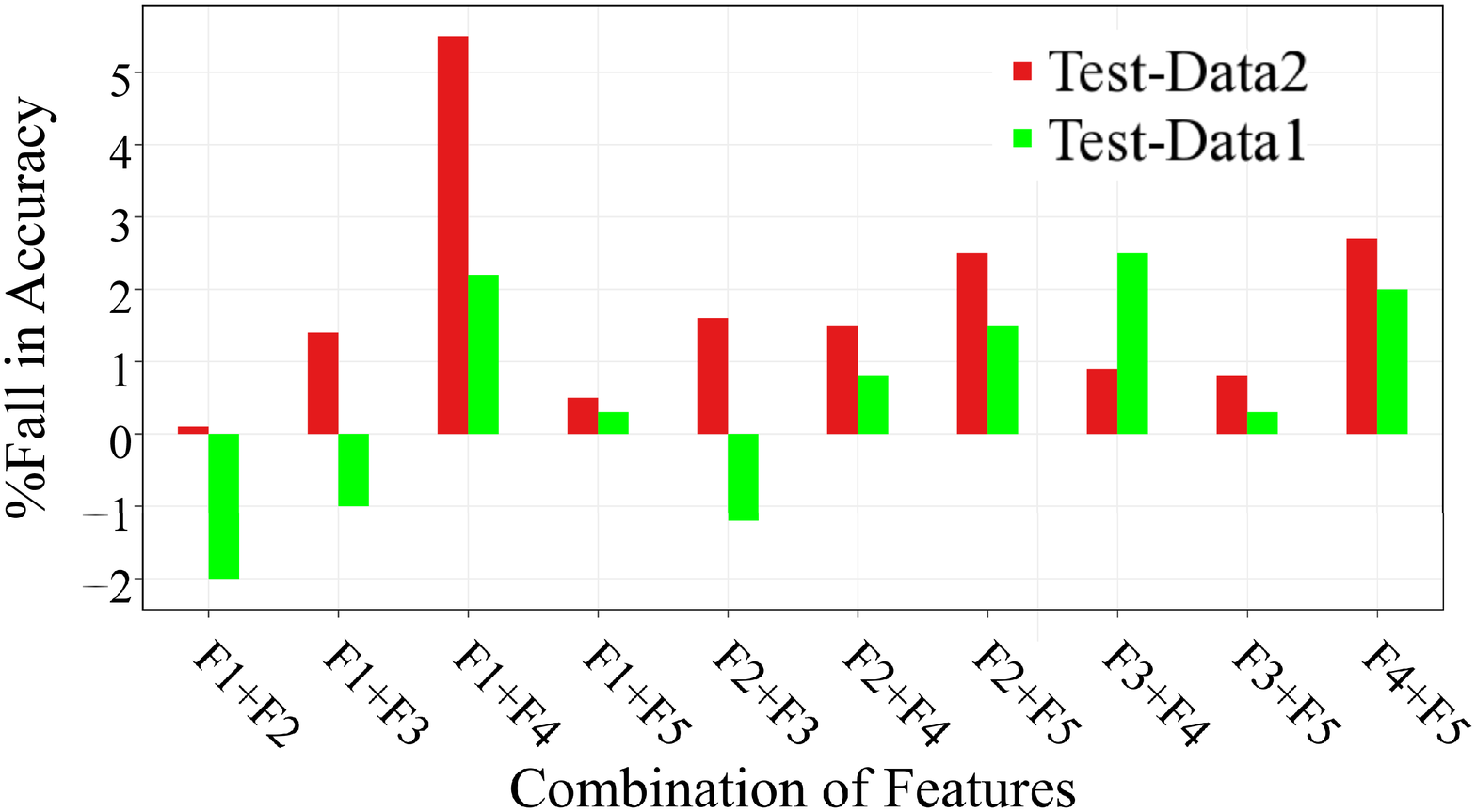}}
\subfigure[\label{fig:leave2out_nb}Naive Bayes]{\includegraphics[width=0.32\textwidth]{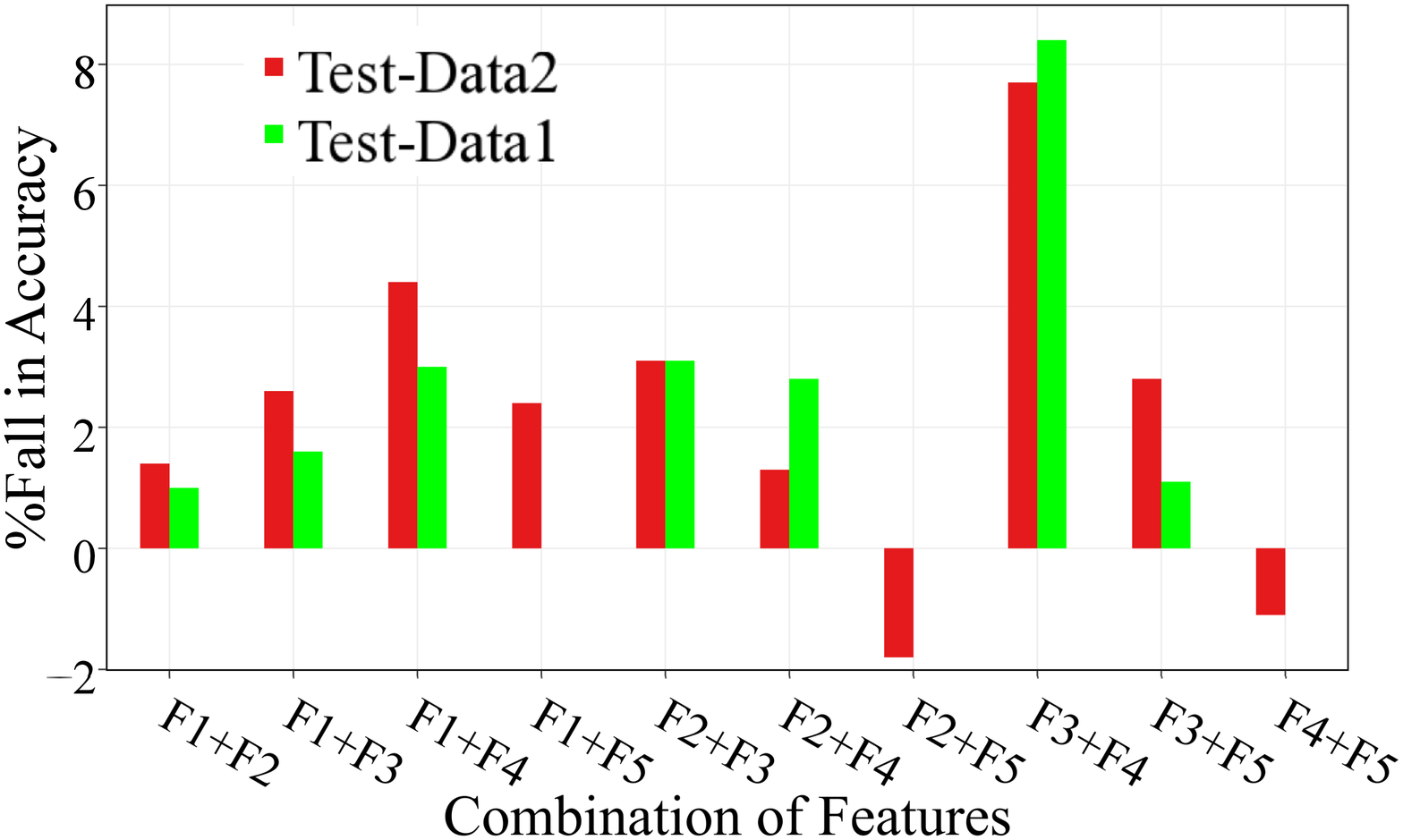}}
\subfigure[\label{fig:leave2out_rf}Random Forest]{\includegraphics[width=0.32\textwidth, height=3.4cm]{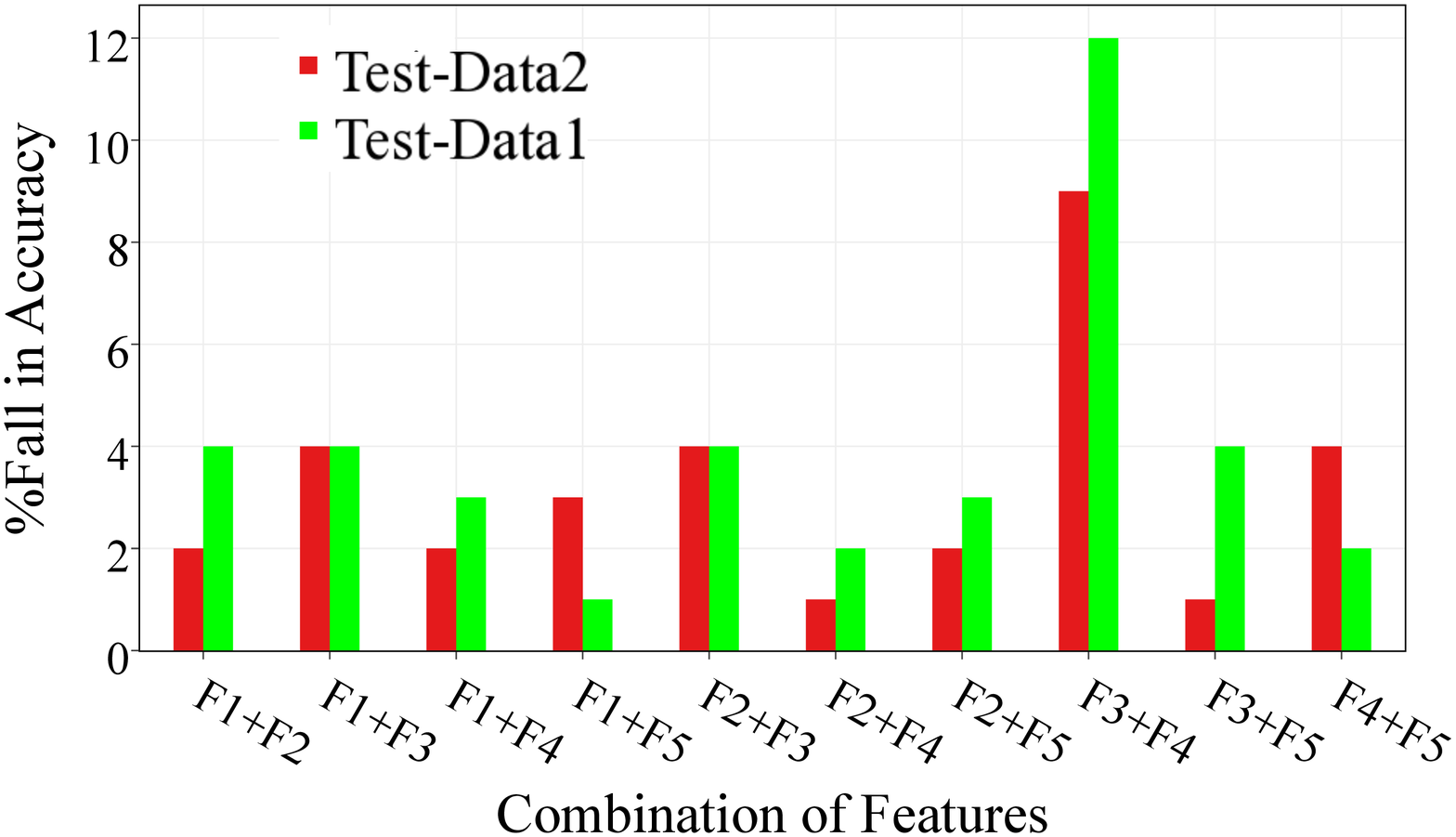}}
}
\caption{\textbf{Percentage Fall in Accuracy of One-Class Classifiers During Leave-P-Out Compilation (P=1- One Feature (Top), P=2- Two Features (Bottom)).} Source: Agarwal et al. \cite{agarwal2016eisic}}\label{fig:leave-p-out}
\end{figure*} 

\textbf{Intent Classification: } An intent of a post (consisting of free-form text) cannot be fully determined only by mining the keywords in the content. But it also requires to understand and predict the psychological tendency, sentiment tones and language of the narrative. It also requires to analyze the semantic role of topic related keywords used in the post. We perform classification on Tumblr posts by training our model on sentiment, semantic and language cues based features of a text. On a high level, we create a vector space of $5$ features set (F1 to F5) which is further categorized into $15$ unique vectors. Table \ref{tab:grouped_features} shows the list of all features extracted and grouped into $5$ feature vectors. We define intent classification as a one-class classification problem. Therefore, our training data contains only positive class (intent) posts. We implement three different one-class classifiers (Random Forest (RF), Naive Bayes (NB) and Decision Tree (DT)) and compare their accuracy for the posts classified as topic in Stage $1$. We train our model for each classifier and perform $5$ fold cross validation. As discussed in Section \ref{sec:experimental_setup} and shown in Figure \ref{fig:stats_infographic}, only $12$\% of the posts are labeled as intent posts making our experimental dataset highly imbalanced. Further, intent classifier takes only the topic posts as an input classified by topic classifier which is again a small subset of whole dataset. Therefore, we select classification algorithms that works for small training data.
\section{Performance Evaluation}
As described in Section \ref{subsec:classification}, proposed method is a cascaded ensemble learning classifier in which topic classifier uses complete experimental dataset as an input while intent classifier takes input from Stage $1$. In this Section, we present the accuracy results of each classifier and also discuss the influence of topic classification's accuracy on intent post classification. Based upon the inter-annotator agreement results, we evaluate the accuracy of our classifier by comparing the observed results against actual labeled class. We conduct our experiments on $2,419$ posts, consistently labeled by both annotators. Proposed topic classifier classifies $346$ posts as target (topic) class and $2,073$ posts as unknown. Table \ref{tab:topic_classification_accuracy} reveals that there is a misclassification of $3.8$\% and $1.6$\% in identifying target and outliers (unknown) posts. Since, the focus of our study is to identify all such posts that have racist or radicalized intent, our aim is to achieve high precision as well as high recall. Our results reveal that for topic classification, we are able to achieve a precision of $73$\% ($253$/($253$+$93$)) and a recall of $86$\% ($253$/($253$+$39$)).\\
\begin{table}[t]
\sffamily
\normalsize
\centering{
\caption{\textbf{Confusion Matrix for Topic Classification}}\label{tab:topic_classification_accuracy}
\begin{tabular}{ll|ll}
\multicolumn{2}{c}{} &\multicolumn{2}{c}{Predicted}\\
\multicolumn{2}{c|}{} &Topic & Unknown\\
\cline{2-4}
\multirow{2}{*}{Actual}& \multirow{1}{*}{Topic} & TP=253 & FN=39\\
& \multirow{1}{*}{Unknown} & FP=93 & TN=2034\\
\end{tabular}
}
\end{table}
\begin{table}[t]
\sffamily
\normalsize
 \centering
 \scriptsize
 \caption{\textbf{Performance Evaluation Metrics for Intent Classification.} Source: Agarwal et al. \cite{agarwal2016eisic}}
 \resizebox{\linewidth}{!}{%
 \begin{tabular}{l|lll|lll}
\multicolumn{1}{c}{}& \multicolumn{3}{c}{Test-Data1} & \multicolumn{3}{c}{Test-Data2}\\
&DT &RF &NB &DT &RF &NB \\
\hline
Recall &0.79&0.82&0.79&0.82&0.84&0.83\\
Precision &0.72&0.78&0.74&0.75&0.81&0.78\\
 \end{tabular}}
 \label{tab:accuracy metrics}
\end{table}
\indent Given that our data is highly imbalanced and only $12$\% of the posts are labeled as target (intent) class, we execute each of our classifiers (RF, NB, and DT) using a $5$ fold cross validation over the experimental dataset. Since, the accuracy measures are biased towards the majority class, we evaluate the performance of intent classifier using two standard information retrieval metrics i.e. precision and Area Under Operator Receiver Curve (AUC). Due to the misclassification in topic modeling, we evaluate the performance of intent classification in two steps. We first execute our model on all $346$ posts (Test-Data$1$) classified as topic in previous stage. In second iteration, we evaluate the performance of intent classifier on $253$ Tumblr posts (Test-Data$2$) correctly classified as topic. Table \ref{tab:accuracy metrics} shows the accuracy metrics for Random Forest (RF), Decision Tree (DT) and Naive Bayes (NB) algorithms.\\
\begin{figure}
 \centering
 \includegraphics[width=0.48\textwidth]{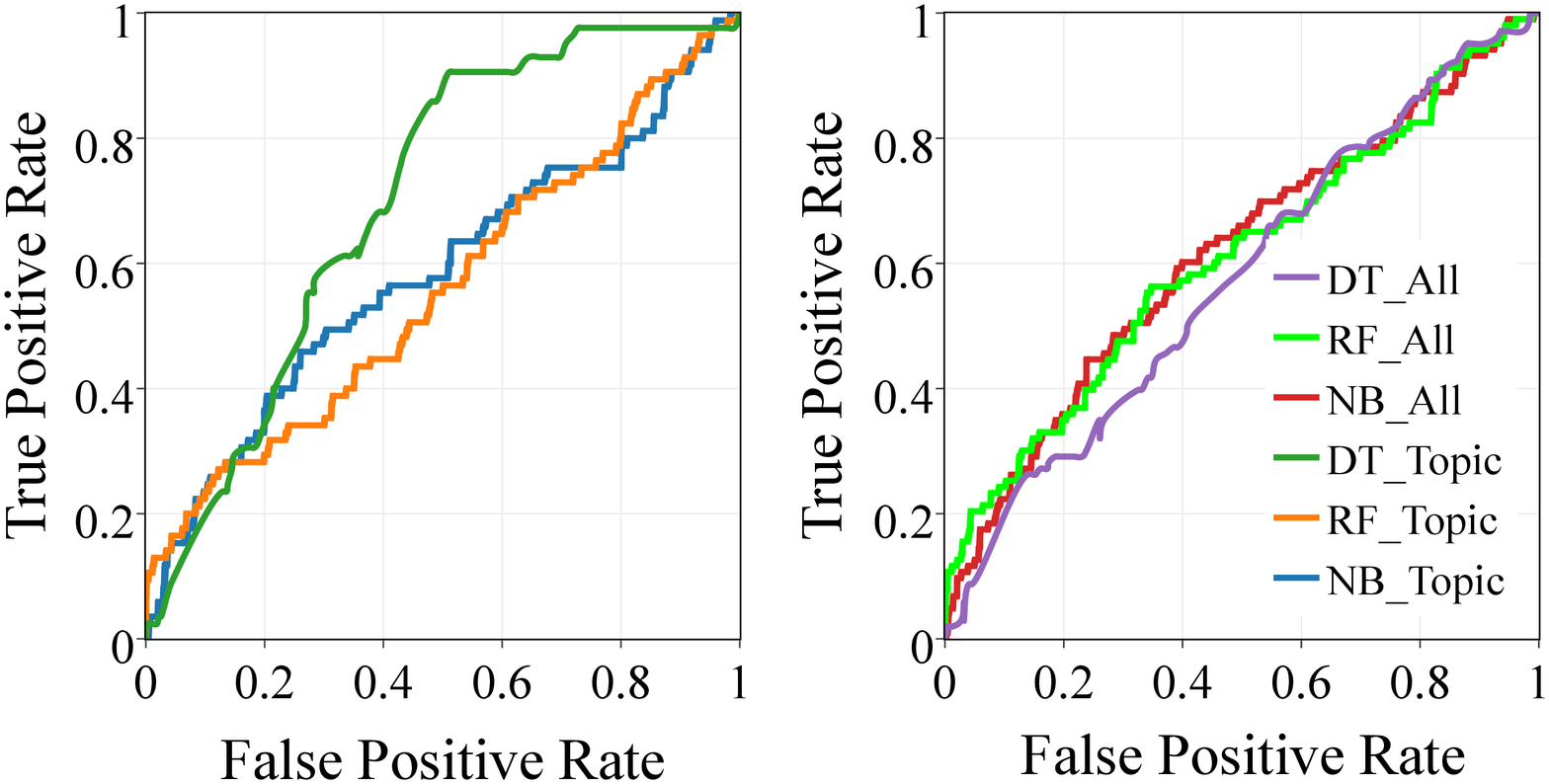}
 \caption{\textbf{ROC Curve for Test-Data1 (Right) and Test-Data2 (Left).} Source: Agarwal et al. \cite{agarwal2016eisic}}
 \label{fig:roc_curve}
\end{figure}

Our results reveal that one-class intent classifier gives higher precision rate for Test-Data$1$ (refer to Table \ref{tab:accuracy metrics}). However, filtering non-topic based posts from the dataset further improves the accuracy of intent classification. This is probably associated with the fact that unknown posts represent a broad range of sentiments and language cues. Table \ref{tab:accuracy metrics} reveals that Random Forest outperforms Naive Bayes and Decision Tree algorithms and gives the maximum precision ($0.78$, $0.81$) and recall ($0.82$, $0.84$) for Test-Data1 and Test-Data2. In fact, both Naive Bayes and Random Forest generate almost similar classification results for topic posts with a difference of $1$\% to $2$\%. Our results reveal that wrongly classified posts at Stage $1$ provokes a decrement in accuracy of intent classification. As shown in Table \ref{tab:accuracy metrics}, classification accuracy for Test-Data$2$ is higher than Test-Data$1$. Figure \ref{fig:roc_curve} shows the ROC curves generated for each type of classifiers executed for both Test-Data1 and Test-Data2. Graphs in Figure \ref{fig:roc_curve} shows that given a set of posts $P=\{P_{i} \mid 1 \le i \le n \mid C(P_{i}) = Topic\}$, Decision Tree based intent classifier has the high probability ($\sim$ $0.7$) to classify them as target class. While, Random Forest and Naive Bayes have almost equal probability ($0.55$) to classify a post as intent or unknown. Figure \ref{fig:roc_curve} reveals that if the taxonomy of a post is unknown (Test-Data1) then each algorithm has a probability of approximately $0.60$ to classify it as intent post.\\

In order to evaluate the impact of each feature on intent classification performance, we test the leave-p-out cross validation for both Test-Data1 and Test-Data2. Figure \ref{fig:leave-p-out} illustrates the percentage of fall in precision of each classifier and for both the datasets. Negative values rather shows the increment in precision. Figure \ref{fig:leave1out_dt} shows that in Test-Data1, removing F2 and F3 individually from the feature space does not impact the overall performance of Decision Tree (\textless $1$\%). While removing writing tone feature i.e. F4 decreases the precision by $4$\%. In fact, for Test-Data2, removing document sentiment vector from the feature space, it increases the performance of Decision Tree by $2$\%. It is possibly due to the reason because emotion tone gives a detailed classification of emotions (anger, fear, joy, sadness and disgust) while document sentiment feature gives overall sentiment of a post that can be biased in longer posts (word length \textgreater $100$). Figure \ref{fig:leave1out_nb} reveals that in Naive Bayes algorithm, removal of any feature from Test-Data2 impacts the performance of classifier with a reasonably high percentage of fall in precision. If we remove feature F1 or F4, it decreases the overall precision upto $3$\%. Similarly, if the taxonomy of a post is unknown (Test-Data1) then removing emotion tone (F3) or language tone (F4) decreases the precision by $2$\%. Similar to Naive Bayes, for Random Forest algorithm (Figure \ref{fig:leave1out_rf}), removal of any feature declines the classifier's performance upto $2$\%. While, for any unknown post, emotion tone ($F3$) and writing cues of the narrative ($F4$) are the most discriminatory features as removal of these features can decrease the performance of algorithm upto $4$\%.\\

We also report the variation in performance of classifiers if a combination of two features is removed from the training model. Leaving out two features at once also reveals the relative influence of each vector in feature space. Figure \ref{fig:leave2out_dt} reveals that feature F3 and F4 are the most discriminatory features as removal of any of these vectors does not influence the performance of other features but we observe a fall in the overall precision rate. For example, removing feature F1 (that increases the precision of Decision Tree algorithm upon leaving out individually) with F4 decreases the precision by $6$\% for Test-Data1 and $2.25$\% for Test-Data2. Similarly, leaving features F2 or F3 along with most of other features (F2 and F4) decreases the performances by $1$\% to $2$\% for both datasets. However, for Test-Data1, leaving these features out along with F1 rather increases the performance. It reveals that in Decision Tree intent classification, Feature F1 negatively impacts the performance of other features. In Naive Bayes intent classification, we find that for Test-Data1, F2 is an important feature for identifying intent posts (Figure \ref{fig:leave2out_nb}). This is possible because if the taxonomy of a post is unknown then semantic tagging of text can be an important feature for identifying the topic related posts. Figure \ref{fig:leave2out_dt} also reveals that in Naive Bayes classifier (Test-Data1), social tone of a text (F5) declines the performance of other features. For example, removing F1 individually decreases the precision by $3$\% while combining it with F5 does not make any change in the accuracy. Similarly, leaving out F3 and F4 features from training model individually makes a fall of $2$\% in overall performance while combining any of them with F5, the accuracy rather improves by $1$\% to $2$\%. It happens because if the posts are not topic related then they might have a wide range of taxonomy which impacts the social tone of a narrative. Due to the sparsity in social tendency attributes, it increases the number of false alarms. Unlike Decision Tree or Naive Bayes algorithms, in Random Forest, removing a combination of any two feature vectors decreases the performance rate of intent classifier for each dataset. Figure \ref{fig:leave2out_rf} reveals that removing any feature along with F3 declines the precision by at least $4$\%. While removing them with F4 can lower the performance by $2$\% to $4$\%. Our results reveal that emotion tone (F3) and writing cues (F4) are the two most discriminatory features for identifying intent post while using any of three classifiers and datasets. Semantic tagging (F2) and social tendency of narratives (F5) are two other important features if the post has a wide range of topics or emotional range making a post ambiguous. Classification results support our hypotheses that sentiment and semantic of text can be used to identify the language cues and personality traits of author and classify the intent post on microblogging platforms.\\

\textbf{Limitations: } In this paper, we conduct our analysis only on English language posts. Our proposed approach has dependencies with the open source APIs used for the feature extraction. If a post contains multi-lingual text (for example, Arabic + English) then the APIs might not be able to extract the taxonomy or semantic features accurately. We make our model generalized and it can be used to identify racist and radicalized intent for any given text. However, the model might require some pre-processing and large training data for microposts as the topic modeling and tone analysis might not be $100$\% accurate for very short text such as tweets.

\section{Conclusions and Future Work}
In this paper, we study the problem of identifying racist and radicalized Tumblr posts based on the intent of narrative. We formulate our problem as a cascaded ensemble learning problem and propose a two-stage one-class classification approach to solve the problem. Our result shows that the proposed approach is effective for identifying intent posts unlike previous keyword based techniques. Our experimental results shows that emotion tone, writing cues and social personality traits of an author are discriminatory features for identifying the intent of the post. Further, topic classification of posts and filtering non-topic based (or noisy) posts improves the performance of the proposed intent classification.\\
\indent Future work includes addressing the limitations of present study and improving the accuracy of linguistic features. Identification of multi-lingual posts by doing a sentence level language detection and enhancing the translated content for identifying intent posts. As mentioned in the previous sections, Tumblr is popularly known for the use of reaction gif images. Therefore, our future work involves mining users' reactions from attached external images and enrichment of linguistic features of a post. Presence of long text in tags gives more information about the intent of an author as well as the content of the post. Future work also includes sentence detection in tags and identifying linguistic features at tag-level.
\bibliographystyle{IEEEtran} 
\bibliography{isi} 

\begin{thebibliography}{10}
\providecommand{\url}[1]{#1}
\csname url@samestyle\endcsname
\providecommand{\newblock}{\relax}
\providecommand{\bibinfo}[2]{#2}
\providecommand{\BIBentrySTDinterwordspacing}{\spaceskip=0pt\relax}
\providecommand{\BIBentryALTinterwordstretchfactor}{4}
\providecommand{\BIBentryALTinterwordspacing}{\spaceskip=\fontdimen2\font plus
\BIBentryALTinterwordstretchfactor\fontdimen3\font minus
  \fontdimen4\font\relax}
\providecommand{\BIBforeignlanguage}[2]{{%
\expandafter\ifx\csname l@#1\endcsname\relax
\typeout{** WARNING: IEEEtran.bst: No hyphenation pattern has been}%
\typeout{** loaded for the language `#1'. Using the pattern for}%
\typeout{** the default language instead.}%
\else
\language=\csname l@#1\endcsname
\fi
#2}}
\providecommand{\BIBdecl}{\relax}
\BIBdecl

\bibitem{10.2307/2265305}
J.~Cohen, ``Freedom of expression,'' \emph{Philosophy \& Public Affairs},
  vol.~22, no.~3, pp. 207--263, 1993.

\bibitem{smith2008language}
A.~G. Smith and P.~e.~a. Suedfeld, ``The language of violence: Distinguishing
  terrorist from nonterrorist groups by thematic content analysis,''
  \emph{Dynamics of Asymmetric Conflict}, vol.~1, no.~2, pp. 142--163, 2008.

\bibitem{agarwal2015open}
S.~Agarwal, A.~Sureka, and V.~Goyal, ``Open source social media analytics for
  intelligence and security informatics applications,'' in \emph{Big Data
  Analytics}.\hskip 1em plus 0.5em minus 0.4em\relax Springer, 2015, pp.
  21--37.

\bibitem{norton2011whites}
M.~Norton and S.~Sommers, ``Whites see racism as a zero-sum game that they are
  now losing,'' \emph{Perspectives on Psychological Science}, 2011.

\bibitem{wang2015mining}
J.~Wang, G.~Cong, and et~al., ``Mining user intents in twitter: A
  semi-supervised approach to inferring intent categories for tweets,'' in
  \emph{AAAI}, 2015.

\bibitem{purohit2015}
H.~Purohit, G.~Dong, and et~al., ``Intent classification of short-text on
  social media,'' in \emph{SocialCom 2015}.

\bibitem{ding2015mining}
X.~Ding, T.~Liu, J.~Duan, and J.-Y. Nie, ``Mining user consumption intention
  from social media using domain adaptive convolutional neural network.'' in
  \emph{AAAI}, 2015, pp. 2389--2395.

\bibitem{geetha2014feature}
P.~Geetha, R.~Chandresh, and et~al., ``Feature selection framework for data
  analytics in microblogs,'' in \emph{'Emerging Research in Computing,
  Information, Communication and Applications' ERCICA}, 2014.

\bibitem{wang2013mining}
J.~Wang, W.~X. Zhao, H.~Wei, H.~Yan, and X.~Li, ``Mining new business
  opportunities: Identifying trend related products by leveraging commercial
  intents from microblogs.'' in \emph{EMNLP}, 2013, pp. 1337--1347.

\bibitem{prentice2011analyzing}
S.~Prentice, P.~J. Taylor, P.~Rayson, A.~Hoskins, and B.~O?Loughlin,
  ``Analyzing the semantic content and persuasive composition of extremist
  media: A case study of texts produced during the gaza conflict,''
  \emph{Information Systems Frontiers}, vol.~13, no.~1, pp. 61--73, 2011.

\bibitem{agarwaltumblr}
S.~Agarwal and A.~Sureka, ``Semantically analyzed metadata of tumblr posts and
  bloggers, mendeley data, v1, http://dx.doi.org/10.17632/hd3b6v659v.1,'' 2016.

\bibitem{agarwal2016eisic}
A.~Swati and S.~Ashish, ``But i did not mean it!- intent classification of
  racist posts on tumblr,'' in \emph{6th IEEE European Intelligence \& Security
  Informatics Conference (EISIC), Uppsala, Sweden}.\hskip 1em plus 0.5em minus
  0.4em\relax IEEE, 2016.

\bibitem{chen2012detecting}
Y.~Chen, Y.~Zhou, S.~Zhu, and H.~Xu, ``Detecting offensive language in social
  media to protect adolescent online safety,'' in \emph{Privacy, Security, Risk
  and Trust (PASSAT), SocialCom}.\hskip 1em plus 0.5em minus 0.4em\relax IEEE,
  2012, pp. 71--80.

\bibitem{agarwal2016spider}
S.~Agarwal and A.~Sureka, ``Spider and the flies: Focused crawling on tumblr to
  detect hate promoting communities,'' \emph{arXiv preprint arXiv:1603.09164},
  2016.

\bibitem{burnap2016us}
P.~Burnap and M.~L. Williams, ``Us and them: identifying cyber hate on twitter
  across multiple protected characteristics,'' \emph{EPJ Data Science}, vol.~5,
  no.~1, p.~1, 2016.

\end{thebibliography}
\end{document}